\documentclass[twocolumn,showpacs,groupedaddress,aps,floatfix]{revtex4}

\usepackage{graphics}

\begin{document}

\title{Integrated optical components on atom chips}

\author{S. Eriksson}
\author{M. Trupke}
\author{H. F. Powell}
\author{D. Sahagun}
\author{C. D. J. Sinclair}
\author{E. A. Curtis}
\author{B. E. Sauer}
\author{E. A. Hinds}
\affiliation{Blackett Laboratory, Imperial College, London SW7 2BW, UK}
\author{Z. Moktadir}
\author{C. O. Gollasch}
\author{M. Kraft}

\affiliation{School of Electronics and Computer Science,
University of Southampton, Southampton SO17 1BJ, UK}

\begin{abstract}
We report on the integration of small-scale optical components
into silicon wafers for use in atom chips. We present an on-chip
fibre-optic atom detection scheme that can probe clouds with small
atom numbers. The fibres can also be used to generate microscopic
dipole traps. We describe our most recent results with optical
microcavities and show that single-atom detection can be realised
on an atom chip. The key components have been fabricated by
etching directly into the atom chip silicon substrate.
\end{abstract}


\maketitle
\section{Introduction}

Atom chips offer a convenient way to miniaturise experiments in
atomic physics~\cite{Hinds1999,Folman2002}. Microstructures on the
surface of the chip produce magnetic and/or electric fields, which
can be used to confine and manipulate cold alkali atoms near the
surface of the chip. Several research groups can now prepare
Bose-Einstein condensates (BECs) in microscopic magnetic traps on
atom chips. The large field gradients that can be generated near a
microstructured surface permit the controlled tunnelling of atoms
over micron or sub-micron lengths.  This makes the atom chip a
natural platform for applications in coherent
matter-wave control such as miniaturised atom
interferometry~\cite{Hinds2001,Hansel2001}, quantum information
processing~\cite{Calarco2000}, and the study of low-dimensional
quantum gases~\cite{1Dgas}. Interaction between the cold atoms
and the room temperature surface of the atom chip can be
detrimental to applications that require the most sensitive
control of the atom. For example, thermal fluctuations of the
charges in the conducting chip surface produce magnetic field
noise. This can relax atomic spin coherences and populations,
thereby destroying the quantum coherence of the
atoms~\cite{Henkel1999,Jones2003,Harber2003,Rekdal2004,Lin2004}.
Furthermore, microscopic imperfections in the materials of the
atom chip can cause roughness in the trapping potential, leading
in extreme cases to fragmentation of the atom
clouds~\cite{Fortagh2002,Leanhardt2003,Jones2004,Esteve2004}.
These phenomena are now well understood, and their consequences
can be minimised by careful consideration of the chip design.  As
an example, very recent results from experiments with atom chips
based on permanently magnetised patterns on videotape show that
the spin-flip-induced loss rate is substantially decreased due to the
reduced thickness of conducting
material on the chip~\cite{Curtis2005_1,Curtis2005_2}. Further improvement is
expected by using even thinner multilayered
films~\cite{Eriksson2004}. Refinement of fabrication techniques
for current carrying wires is also in progress.

With this new understanding and control over atom-surface
interactions it is becoming possible to manipulate the quantum
states of a few atoms in very tight traps only a few micrometres from
the surface of an atom chip.  This raises the crucial question of
how to detect a few atoms close to the chip. Hitherto, the
standard method for probing clouds of trapped atoms has been by
absorption imaging or fluorescence detection. The laser light is
typically delivered to the chip by external mirrors and lenses.
This technique becomes increasingly difficult to utilise as the
atom-surface distance becomes small and when the cloud contains
very few atoms. Moreover, it is difficult to address individual
atoms in this way. Any  attempt at constructing a large-scale
quantum information processor based on trapped neutral atoms would
greatly benefit from a simple detection scheme able to sense a
single atom and having high spatial resolution. It is therefore
desirable to devise new on-chip detection schemes.

In this paper, we report on our recent efforts towards integrating
micro-optical mirrors and fibres into atom chips based on silicon.
These components enable us to tailor light fields on the micrometre
size-scale, which is commensurate with the relevant trap sizes.
Such light fields are well suited for probing  
a small part of a large atom cloud, or alternatively, for sensing small 
groups of atoms or single atoms. Some of these components are manufactured
as an integral part of the silicon wafer surface and can be used
to detect atoms within a few microns from the surface. Control of
light fields on this scale also opens up entirely new
possibilities for manipulating atoms in microscopic traps.

We have chosen to work with atom chips fabricated on silicon
wafers because this material is well understood and many
fabrication techniques are already well established. We have been
working on a number of devices that use patterns etched into the
silicon wafer. Here, we introduce two of them. (i) An atom
detector based on a pair of optical fibres mounted in a v-groove
etched on the chip wafer. Each fibre can be used to shine light
into the other across a small gap where atoms may be placed. A
few (5\,-\,10) atoms in the gap can then be detected, either through
the phase shift of the light if it is off-resonant, or through the
absorption of near-resonant light~\cite{Horak}. The fibres can
also be used to generate a one-dimensional standing wave pattern,
making an optical lattice of $\sim 100$ anti-nodes in which the atoms
can be manipulated by the optical dipole force. (ii) We describe a
plano-concave optical microcavity where the concave mirror is
formed on an etched spherical indentation in the silicon wafer
itself. The other mirror is attached to the end of an optical
fibre, which acts as the input/output port for light. This
permits on-chip single atom detection.

In the next section we describe a typical silicon atom chip into
which these structures are being integrated. Section
\ref{section:V-grooves} deals with incorporating optical fibres
for on-chip detection of small atom clouds. In section
\ref{section:microcavity} we describe optical microcavities that
are suitable for single atom detection. We conclude with a
summary in section~\ref{section:conclusions}.

\section{The magnetic trap atom chip}

The silicon chip that we are currently using manipulates the atoms
with magnetic fields generated by current carrying wires. The
silicon substrate is covered with a 0.6\,$\mu$m thick insulating
layer of SiO$_2$.  This in turn is covered by a thin Cr adhesion layer and
a 5.5\,$\mu$m layer of Au, both deposited by sputtering.  This
provides a reflective surface for laser light that is used to
collect and cool $^{87}$Rb atoms in a magneto-optical trap (MOT)
in a standard mirror-MOT arrangement.  The current-carrying wires are defined
lithographically by etching narrow trenches in the Au and Cr
layers, a fabrication process that has been described in detail
elsewhere~\cite{Koukharenko2004}.

\begin{figure}
    \begin{center}
        \resizebox{0.75\columnwidth}{!}{
            \includegraphics{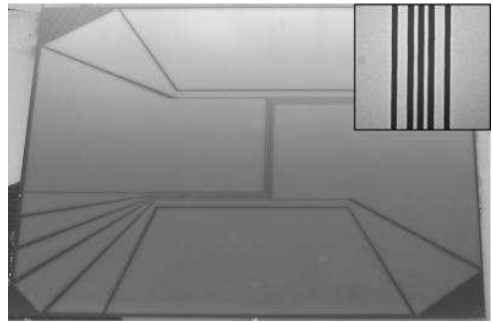}
                        }
    \end{center}
    \caption{The magnetic trap atom chip with dimensions 26.4\,mm (length)
	 by 22.5\,mm (width). The inset shows a high
        resolution microscope image of the central region
        of the chip where the atom cloud is confined. }
    \label{fig:chip}       
\end{figure}

An image of the chip taken before it was mounted in the vacuum
chamber can be seen in Fig.~\ref{fig:chip}. On the chip, four
independent wires trace out a Z-shaped patter. When current flows
through the wire in the presence of a uniform bias field a 
Ioffe-Pritchard trap for weak field seeking atoms is formed in the 
middle of the 7\,mm long central section of the Z~\cite{Reichel1999}.
A microscope image of the central
section is shown in the inset of Fig.~\ref{fig:chip}. In this part
of the chip, there are four parallel wires. The two inner wires
are 33\,$\mu$m wide, while the outer pair have a width of
85\,$\mu$m. The centre-to-centre distances are 83\,$\mu$m and
300\,$\mu$m for the thin and thick wire pairs respectively.
The chip has two additional `end-wires' which run orthogonal to
the centre section. These wires can be used to provide additional
trap depth, or alternatively to shift the magnetic trap along the
Z-wires.

The chip is mounted on a stainless steel base plate, which is in
good thermal contact with the vacuum flange. The high thermal
conductivity of silicon is a useful feature of this particular chip
(for a review of the properties of
common materials that are used to manufacture atom chips, the
reader should consult Ref.~\cite{Reichel2002}).
In ultra-high vacuum, the two inner wires can each carry currents
up to 1.7~A for one second with only a few degrees increase in 
temperature. The corresponding
current limit for the outer Z-wires is 2.5~A, whilst for the
end-wires it is 8.0~A. With all wires at these currents, a
magnetic trap 1 mm from the chip surface has a
trap depth of approximately 1\,mK  which
is sufficient to hold a laser cooled cloud produced in the MOT.

All six wires on the chip can be independently controlled. With
the appropriate choice of wire currents and bias fields, the atom
cloud can be moved around on the chip and split in various ways.
This kind of control over the cloud is essential for many of the
experiments that will be performed with the optical components
described in the next sections of this paper.

\section{Fibre-optic detection of atoms}
\label{section:V-grooves}

\begin{figure}
    \begin{center}
        \resizebox{0.90\columnwidth}{!}{%
            \includegraphics{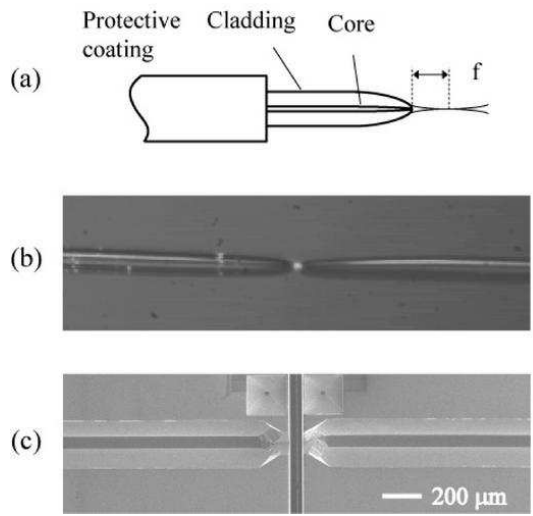}
                        }
    \end{center}
    \caption{Tapered optical fibres. (a) Diagram of the fibre end, (b)
        Microscope image of two opposing fibres with light
        propagating between them. (c) SEM micrograph of
        silicon v-grooves for mounting fibres in an atom chip. The
	scale is the same in (b) and (c).}
    \label{fig:v-grooves}       
\end{figure}

In principle, fluorescence could be used to detect small numbers
of atoms, but it is difficult to achieve a high collection
efficiency and each scattering event heats the atom, whether or
not the scattered photon is detected. The alternative is to use a
directed light beam, which can be efficiently collected and used to
measure the absorption or phase shift due to the atoms.
Signal-to-noise arguments show that the latter method is typically
preferable, provided the cross-sectional area of the light beam is
made small enough~\cite{Horak}. A suitably tight waist can be
achieved using a tapered optical fibre, as illustrated in
Fig.~\ref{fig:v-grooves}\,(a), which has a focal spot at a 
working distance $f$ from the tip of the fibre. 
Over most of its length, the fibre we use is a standard 780~nm
single-mode fibre with a protective acrylate coating and an
outside diameter of 250\,$\mu$m. A few millimeters from the end,
the protective coating is removed to reveal the 125\,$\mu$m
diameter cladding. A few hundred micrometers from the end, both
the core and the cladding begin to taper, and the fibre terminates
in a curved surface. This leads to a focal spot with an
intensity distribution whose full width
at half maximum  is $w$ at the working distance $f$ from
the fibre tip. We have been working with fibres where $f\simeq
25\mu$m and $w\simeq 2.8~\mu$m.

A single tapered fibre may be used to sense atoms by fluorescence,
provided that they are held sufficiently near the fibre tip. This
could be achieved, for example, by sending 795\,nm light through
the fibre to form an optical dipole force trap, tuned to the red
side of the $^{87}$Rb D1 (5$^2$S$_{1/2}$ - 5$^2$P$_{1/2}$)
transition. $^{87}$Rb atoms held by a magnetic trap elsewhere on
the chip are moved into the detection zone near the focal spot of
a fibre and are captured in this trap. Resonant light tuned to the
D2 transition at 780 nm is then passed through the same fibre in
order to induce fluorescence, some of which is collected by the
fibre and appears as reflected light. Unfortunately, even in the
absence of atoms there is a background of reflected light which is
difficult to reduce below the level of $\sim 0.1 \%$ of the input.
With careful filtering of the reflected trapping light this method
has the capability to detect several tens of atoms.

A much better scheme is to use two fibres, mounted face-to-face.
Fig.~\ref{fig:v-grooves}~(b) shows a microscope image of the
arrangement, 
with light propagating between two tapered fibres.
After adjusting the focal spots to coincide we are
able to obtain fibre to fibre coupling efficiencies of up to  87 \%, 
more than enough for detection by absorption to become
feasible. The situation is now very similar to standard absorption
imaging, except that detection takes place in a small,
well-localised part of the atom chip.
A second important difference is that every photon in
the light beam passes through a waist of small area $\pi
w^2/4$\,$\simeq$\,6\,$\mu$m$^2$. The resonant absorption cross section
$3 \lambda ^2/2\pi$ is roughly 20 times smaller than this, resulting
in a $\sim 5\%$ absorption for each atom in the waist of the light
beam. This method is suitable for detecting 5-10 atoms. An
alternative version of this detector is to use non-resonant light
and to measure the optical phase shift due to the non-resonant
interaction.  This has a similar limit of detection sensitivity,
as discussed in Ref.~\cite{Horak}.

In order to establish good mechanical stability and alignment of
the fibres on the atom chip, we etch v-grooves into the silicon
substrate. Fig.~\ref{fig:v-grooves}~(c) shows a scanning electron
microscope (SEM) micrograph looking down on them. The length of
each groove can be several centimetres and its width is chosen to fit the
optical fibre stripped down to its cladding.
Fig.~\ref{fig:v-grooves}~(c) shows the central 2\,mm, which on
this particular chip includes a trench for an atom guide wire (the
dark vertical rectangular feature). Near the centre, where the
fibres come to within a few tens of micrometers of one another,
the groove becomes narrower to support the tapered fibre ends. The
grooves are manufactured by depositing masking layers of 40\,nm of
SiO$_2$ and 160\,nm of SiN on a clean silicon wafer. The surface
is photolithographically patterned to make openings that have a
rectangular shape with constant width for most of the length. Near
the fibre tips, the width of the opening in the mask is stepped
down to match the narrower tapered ends of the fibers. The masking layer
is removed in these areas by dry etching, leaving windows through
which the silicon is then wet etched, using 30\,\% KOH by volume at
70\,$^\circ$C. The etching time is optimised to create a groove of
the desired depth.

This structure of two fibres in a v-groove can also be used to
generate optical lattices in one dimension by guiding light
through both fibres. We propose to use this geometry to create a
Mott-insulator transition~\cite{Bloch} on an atom chip. The basic
idea is to increase the optical lattice intensity adiabatically
from zero in the presence of a BEC between the fibre ends. A key
point here is that the small transverse dimension of the light
beam allows the volume of each well in the lattice to be small.
This ensures an atom-atom repulsion energy in the kHz domain,
allowing the transition to be achieved adiabatically on the millisecond
timescale. Once the cloud has made the transition into a Mott
state, the fibres can be used for detection of the atom ordering
by coherent Bragg scattering from the lattice.

\section{Optical microcavities}
\label{section:microcavity}
\begin{figure}
\vspace{0.5cm}
    \begin{center}
        \resizebox{\columnwidth}{!}{%
            \includegraphics{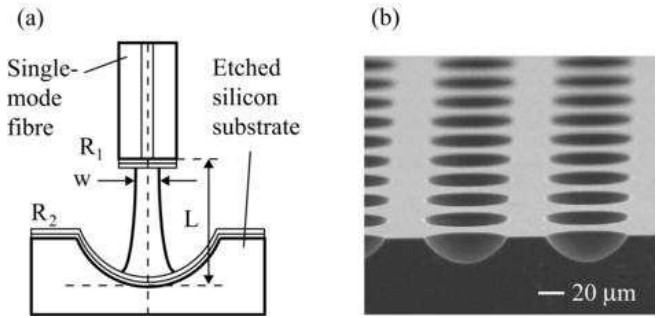}
                        }
    \end{center}
    \caption{The optical microcavity.
        (a) Plano-concave optical microcavity with length $L$
        and cavity waist $w$. The plane mirror with reflectivity $R_1$ is formed on
        the fibre tip by applying a pull-off coating. The concave mirror
        with reflectivity $R_2$ is formed by sputtering Au onto the etched silicon wafer
        (see main text for details). (b) SEM micrograph of an array of
        isotropically etched spherical mirrors on a silicon wafer.}
    \label{fig:cavitydiagram}       
\end{figure}

In order to bring the optical detection sensitivity down to one
atom, the light beam needs to pass several times back and forth
through the region where the atom is located. In other words, we
require an optical cavity. For the purpose of detecting one atom
it is not necessary to have a cavity with particularly high finesse 
provided the
waist size is  small. In that case, the main challenge lies
in constructing cavities that can be integrated on the atom chip.
To this end, we have been working on plano-concave microcavities
with the curved mirror etched into the silicon wafer. The plane
mirror is a flat, cleaved fibre tip with a high reflectivity coating
applied to it. The general scheme of the cavity is illustrated in
Fig.~\ref{fig:cavitydiagram}~(a). The concave mirror of the cavity
is fabricated by means of isotropic etching, a widely used
technique in micromachining in which the silicon wafer is etched
at approximately the same rate in all directions. The 
etching solution is continuously stirred
during the etching process, resulting
in a surface that is approximately spherical. The etchant is a
mixture of hydrofluoric acid (H) and nitric acid (N), diluted in
acetic (A) acid. The surface morphology and  etch rate produced by
these chemicals are highly dependent on the concentration of each
component, and on the agitation. The relative amounts we prefer to
use are H:N:A = 9:75:30 by volume. Further details of the fabrication
process can be found elsewhere~\cite{Moktadir2004}. An SEM
micrograph of a typical array of spherical mirror templates is
shown in Fig.~\ref{fig:cavitydiagram}~(b). In order to investigate
cavities of various lengths, we have adjusted the etching
parameters to produce wafers with a range of mirror radii between
50 and 250\,$\mu$m. The etched surface typically has an RMS
roughness of 5\,nm, as measured by atomic force microscopy, often
with some areas noticeably better than others. For our first
measurements, we sputtered an adhesion layer of Cr onto these
substrates, followed by 100 nm of Au. After sputtering, the
roughness increased to $\sim$ 10 nm.

The plane mirror of the cavity, which acts as the input coupler,
is made by gluing a dielectric multilayer transfer coating onto
the plane cleaved end of a single-mode 780\,nm optical fibre (mode
size 2.7\,$\mu$m).  We have chosen a coating with a reflectivity
of $R_1$\,=~98-99\% for light in the near infrared range of the
spectrum in order to match the reflectivity of gold in this
wavelength range. The coating layer is transferred from a glass
plate (to which it is loosely attached by adhesion with a dense
saline solution) by applying an index-matched optical epoxy to the
fibre tip. After the epoxy has set, the tip is pulled away from
the glass plate, breaking the coating around the tip edges.

\begin{figure}[h]
    \begin{center}
        \resizebox{0.90\columnwidth}{!}{%
            \includegraphics{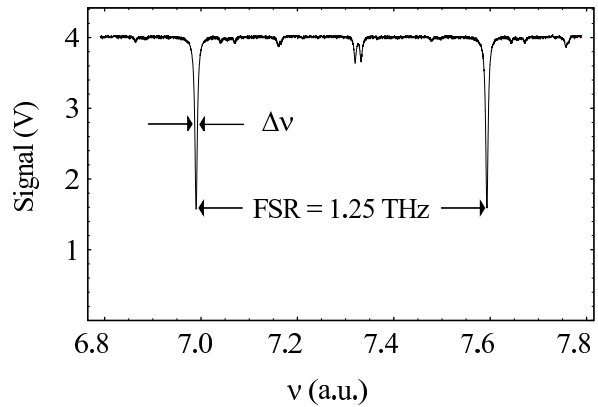}
                        }
    \end{center}
    \caption{Reflected intensity from an optical microcavity with gold-coated spherical mirror. The cavity
        length is scanned over one FSR.}
    \label{fig:cavityfringes}       
\end{figure}

In order to investigate the optical properties of the cavity,
780\,nm light from a diode laser is coupled into the fibre, which
is aimed at a gold-coated mirror. The silicon wafer is translated
longitudinally by a piezoelectric stack to produce a sweep that
can cover several cavity fringes and the reflected intensity is
monitored. Fig.~\ref{fig:cavityfringes} shows the reflected light
signal as a cavity of length  $L$\,=\,120\,$\mu$m is scanned over
one free spectral range (FSR).  In this experiment the curved
mirror had a 180\,$\mu$m radius of curvature and the finesse
(FSR/$\Delta\nu$) was 102.

Taking the fringe visibility together with the value of the
finesse, we calculate that the (intensity) reflectivities of the
mirrors are $R_1=0.989$ and $R_2=0.950$. The
value of $R_1$ corresponds closely to the value specified  by the
manufacturer of the transfer coating. We attribute the low value
of $R_2$ to the surface roughness of the convex mirror. Even so,
this cavity is already good enough to detect single atoms. When
weakly illuminated with light, we calculate that the amplitude of
the reflection dips decreases by a factor of 5 if an atom is
introduced into the cavity~\cite{Rempe}.

For future applications in quantum information processing, it
would be useful to achieve strong coupling between the atom and
the optical microcavity. With this in mind we have started to
investigate how high the finesse of these cavities can be made. As
a first step in this direction we have put a dielectric coating on
one of the silicon wafers and have used a higher-reflectivity
transfer coating on the fibre. First results show a finesse in
excess of 3,000 on selected mirrors, opening the way to new cavity
QED experiments on a chip.

Further work is necessary to incorporate the cavity into an
operating atom chip so that it can be loaded with atoms. 
In order to fine tune the cavity on the
chip we have been developing a micro-actuator, which allows the
translation of the concave mirror in all three directions of
space~\cite{Gollasch2005}. The translation range of the cavity
mirror in the chip plane is large enough to compensate for any
mismatch of the optical axes of the cavity mirrors. 
The cavity length can be tuned to sweep
over several cavity fringes.

\section{Summary}
\label{section:conclusions}

We have presented optical micro-components that can be integrated
into atom chips. We have investigated the properties of a
fibre-optic detection scheme and shown that clouds with small atom
numbers can be probed. We have also developed a high-finesse
optical micro-resonator and have shown that it is suitable for
single atom detection on a chip, coupled to the external world by
a fibre. Silicon is the substrate material that we have used to
develop these new optical components because that is the basis of
some working atom chips. However, our techniques are probably
applicable to any substrate that can be etched or micro-machined.
The challenge from the micro-fabrication side of the work lies in the 
assembly of the various micromachined components of the atom chip, such 
as the three-axis actuator, the gold wires, the detection cavities and 
the v-grooves for the optical fibres. This will require bonding of several 
silicon wafers with good alignment (of the order of a few micrometres). 
Work is in progress to produce an atom chip that
incorporates these various components.
 
\begin{acknowledgments}
The authors would like to thank  Jon Dyne for expert technical assistance.
This work is supported by the UK Engineering and Physical Sciences Research
Council, the Royal Society, and by the FASTNET and QGATES networks of the 
European Union.
\end{acknowledgments}


\end{document}